\documentclass[10pt,onecolumn,aps,prd,preprintnumbers,showpacs,superscriptaddress,nofootinbib,amsmath,amssymb,floats,floatfix,showkeys,notitlepage,longbibliography]{revtex4-2}
\usepackage{orcidlink}
\UseRawInputEncoding
\usepackage{comment}
\usepackage{lipsum}
\usepackage{graphicx}
\usepackage{subfigure}
\UseRawInputEncoding
\usepackage{palatino}
\usepackage{sans}
\usepackage{hyperref}
\hypersetup{colorlinks=true,linkcolor=blue,urlcolor=blue,citecolor=blue}
\usepackage[toc,page]{appendix}
\usepackage[normalem]{ulem}
\usepackage{adjustbox}
\usepackage{latexsym}
\usepackage{amsmath}
\usepackage{amssymb}
\usepackage{amsfonts}
\usepackage{commath}
\usepackage{physics}
\usepackage{dcolumn}
\usepackage{bm}
\usepackage{tikz}
\usetikzlibrary{decorations.pathmorphing}
\usepackage{bigints}
\usepackage{array,tabularx,multirow,booktabs}
\usepackage[tracking=true]{microtype}
\usepackage{soul} %for highlighting
\SetTracking{}{500}
\SetTracking{encoding={*}, shape=sc}{40}
\UseRawInputEncoding %for inputenc error%
\allowdisplaybreaks

\usepackage[utf8]{inputenc}
\usepackage{xcolor} % For colored text if needed

\begin{document} \sloppy
\title{Near extremal RN-AdS control of holographic Josephson transport}

\author{Ali \"Ovg\"un \orcidlink{0000-0002-9889-342X}}
\email{ali.ovgun@emu.edu.tr}
\affiliation{Physics Department, Eastern Mediterranean University, Famagusta, 99628 North
Cyprus via Mersin 10, Turkiye.}

\author{Reggie C. Pantig \orcidlink{0000-0002-3101-8591}} 
\email{rcpantig@mapua.edu.ph}
\affiliation{Physics Department, School of Foundational Studies and Education, Map\'ua University, 658 Muralla St., Intramuros, Manila 1002  Philippines.}

%\date{\today}

\begin{abstract}
We formulate a holographic weak-link construction in which Josephson transport is controlled by the charge sector of a Reissner--Nordstrom-AdS black brane. The model is an Einstein-Maxwell-charged-scalar theory in asymptotically AdS$_4$, with a spatially inhomogeneous boundary chemical potential that creates two superconducting banks separated by a normal or weakly superconducting barrier. The Josephson phase difference is defined from gauge-invariant boundary observable of the charged condensate, rather than from the black hole charge itself, allowing a controlled extension of the standard holographic SNS junction to charged AdS backgrounds. We identify the current-phase relation, critical current, coherence length, and small-phase stiffness as the main observables. In the SNS regime, the critical current and midpoint condensate probe the same proximity scale, while higher harmonics in the current-phase relation diagnose enhanced transparency or departure from the opaque weak-link limit. The new mechanism is the near-extremal RN-AdS throat: as extremality is approached, the emergent AdS$_2\times\mathbb{R}^2$ region can radially modify the Josephson coupling before it reaches the ultraviolet boundary. After removing the ordinary spatial suppression due to the junction width, the residual critical current and phase stiffness are expected to exhibit scaling governed by the infrared dimension of the charged scalar in AdS$_2$. This separates ordinary proximity suppression, smooth finite-density corrections, and genuinely near-extremal throat control, providing a framework for phase-sensitive transport in charged holographic matter.
\end{abstract}

\keywords{Holographic Josephson junctions;
Reissner--Nordstrom-AdS black branes;
Holographic superconductors;
Near-extremal AdS$_2$ throats;
Infrared scaling;
Critical current}

\maketitle

\section{Introduction}
Gauge/gravity duality provides a nonperturbative relation between certain large-$N$ quantum field theories and classical gravitational dynamics in asymptotically anti-de Sitter spacetimes \cite{Maldacena:1997re,Gubser:1998bc,Witten:1998qj}. 
Its usefulness for strongly coupled quantum matter stems from the fact that finite temperature, finite charge density, and real-time response functions can be encoded geometrically in black hole backgrounds and bulk perturbations \cite{Hartnoll:2009sz}. 
Among the most influential applications of this idea is the holographic superconductor, in which a charged scalar field coupled to a Maxwell field in AdS condenses below a critical temperature and gives the boundary theory a superconducting order parameter \cite{Gubser:2008px,Hartnoll:2008vx,Hartnoll:2008kx}. 
The essential mechanism is that the electric field of a charged or finite-density black hole background lowers the effective mass of a charged scalar near the horizon, allowing a source-free condensate to form without destabilizing the ultraviolet AdS boundary theory \cite{Gubser:2008px}. 
In the dual description, the near-boundary coefficients of the scalar and Maxwell fields determine the condensate, chemical potential, charge density, and current response of the boundary system \cite{Hartnoll:2008vx,Hartnoll:2008kx}.

The holographic-superconductor framework has also been generalized in many directions beyond the minimal Einstein--Maxwell--scalar model.  Higher-curvature, Weyl, quasi-topological and higher-derivative corrections were shown to modify the onset of condensation, the optical conductivity and the superconducting gap, thereby providing useful probes of how ultraviolet gravitational interactions affect boundary superconducting transport
\cite{Wu:2010vr,Kuang:2010jc,Kuang:2016edj,Kuang:2013oqa,Kuang:2011dy,Momeni:2014efa,Barrientos:2025rde}.  
Nonlinear electromagnetic sectors, including Born--Infeld, power-Maxwell and arcsin-electrodynamics models, have likewise been used to test the robustness of the condensate and conductivity against nonlinear charge dynamics
\cite{Gangopadhyay:2012am,Gangopadhyay:2012np,Sheykhi:2016kqh,Sheykhi:2017tzb,Kruglov:2018jee}.  
Analytic and hydrodynamic aspects of holographic superconductors and superfluids have clarified the near-critical behavior, sound modes, superfluid transport and finite-density response
\cite{Herzog:2010vz,Herzog:2009md,Herzog:2011ec,Herzog:2012kx}.  
Related developments involving large-dimensional gravity, Riemann-problem dynamics, entanglement entropy and complexity further show that holographic superconducting systems can be embedded in a broader class of gravitational and information-theoretic constructions
\cite{Emparan:2020inr,Herzog:2016hob,Momeni:2015iea,Paul:2025gpk,Momeni:2010jf}.  
These works demonstrate that superconducting transport in holography is highly sensitive to the gravitational background, gauge-sector dynamics and infrared geometry, motivating the present study of how the near-extremal RN--AdS charge sector can modify Josephson observables through the emergent AdS$_2\times\mathbb{R}^2$ throat.

The Josephson effect is a phase-sensitive manifestation of superconductivity in which two superconducting regions separated by a weak link support a nondissipative current controlled by the gauge-invariant phase difference of their condensates \cite{Josephson:1962zz,Likharev:1979zz,Barone_1982}. 
Because the effect depends directly on phase coherence rather than only on the magnitude of the order parameter, it is a natural diagnostic of superconducting stiffness, weak-link transparency, and proximity physics \cite{Likharev:1979zz,Barone_1982}. 
A gravitational dual of an SNS Josephson junction was constructed by imposing a spatially inhomogeneous boundary chemical potential in the minimal holographic superconductor model \cite{Horowitz:2011dz,Takeuchi:2013kra}. 
That construction reproduced the sinusoidal current-phase relation, the exponential suppression of the maximum current with junction width, and the expected dependence of the critical current on temperature \cite{Horowitz:2011dz}. 
Consequently, the problem addressed here is not the first construction of a holographic Josephson junction, but the identification of how black hole charge and near-extremal infrared geometry alter the phase-sensitive transport observable of such a junction.

A bare Reissner--Nordstrom black hole is not, by itself, a Josephson system. 
It carries charge and has an electrostatic potential, but it does not contain two superconducting domains, a weak link, or a gauge-invariant condensate phase difference. 
The appropriate setting is instead a charged holographic superconductor in which the Reissner--Nordstrom-AdS geometry supplies the finite-density normal state, while the Abelian-Higgs sector supplies the charged condensate that breaks the boundary $U(1)$ symmetry \cite{Gubser:2008px,Hartnoll:2008vx,Hartnoll:2008kx}. 
This distinction is important because the Josephson current must be defined through boundary superconducting observable, not through the electric charge of the black hole alone. 
The Reissner--Nordstrom-AdS charge parameter then becomes a control variable for the superconducting weak link rather than a substitute for the weak link itself.

Near extremality, the planar Reissner--Nordstrom-AdS black brane develops a long AdS$_2\times \mathbb{R}^2$ throat that governs the low-energy infrared response of the finite-density boundary theory \cite{Faulkner:2009wj}. 
In such geometries, radial propagation through the throat can modify ultraviolet boundary observables by infrared scaling dimensions associated with the emergent AdS$_2$ region \cite{Faulkner:2009wj}. 
This suggests that the Josephson coupling between two boundary superconducting domains may be sensitive not only to the width and depth of the weak link, but also to the throat length, horizon electric field, and effective infrared dimension of the charged scalar. 
The central question of this work is therefore how the Reissner--Nordstrom-AdS charge sector modifies the current-phase relation, critical current, coherence length, and phase stiffness of a holographic Josephson junction as the system is driven toward near extremality. 

We consider an Einstein-Maxwell-charged-scalar model in asymptotically AdS$_4$ and impose a spatially varying boundary chemical potential that produces two superconducting regions separated by a normal or weakly superconducting barrier. 
The scalar is written in amplitude-phase form, and the dynamics is expressed in terms of gauge-invariant combinations of the Maxwell field and condensate phase, so that the phase difference across the junction is a boundary observable rather than a gauge convention. 
The main observable is the stationary current $J$ as a function of the gauge-invariant phase difference $\gamma$, together with the associated critical current $J_c$, effective coherence length, and small-phase stiffness. 
By comparing the Schwarzschild-AdS-like regime, the finite-density non-extremal RN-AdS regime, and the near-extremal RN-AdS throat regime, we isolate which features reproduce the known holographic Josephson phenomenology and which features are genuinely controlled by the charged near-horizon geometry. 

The paper is organized as follows. 
Section~2 introduces the Einstein-Maxwell-scalar model, the RN-AdS black brane observable, the charged-scalar instability, and the boundary dictionary for the condensate and current. 
Section~3 constructs the inhomogeneous weak link on a charged AdS background, defines the gauge-invariant phase variables, and specifies the boundary observable used to extract the phase difference. 
Section~4 develops the Josephson observables: the current-phase relation, critical current, coherence length, and phase stiffness. 
Section~5 analyzes the near-extremal AdS$_2\times\mathbb{R}^2$ throat and formulates the infrared scaling interpretation of the Josephson coupling. 
Section~6 summarizes the results and outlines extensions involving backreaction, AC response, magnetic interference, alternative quantizations, and the relation to interior condensate dynamics. 
We work in four bulk dimensions with a $(2+1)$-dimensional boundary theory, use units in which the AdS radius may be set to unity after dimensionless ratios have been identified, and denote bulk indices by Greek letters and boundary indices by Latin letters.

\section{Charged holographic superconductors from RN-AdS observable}
\subsection{Einstein-Maxwell-scalar model}
We begin with the minimal bulk model in which a charged scalar can condense in an asymptotically AdS black hole background.  The four-dimensional action is taken to be \cite{Gubser:2008px,Hartnoll:2008vx,Hartnoll:2008kx,Horowitz:2011dz}
\begin{equation}
S=\frac{1}{2\kappa_4^2}\int d^4x\,\sqrt{-g}
\left[
R+\frac{6}{L^2}
-\frac{1}{4}F_{\mu\nu}F^{\mu\nu}
-|D\Psi|^2
-m^2|\Psi|^2
\right] .
\label{2.1}
\end{equation}
Here $L$ is the AdS radius, $\kappa_4^2=8\pi G_4$, and the Maxwell field and charged scalar are defined by
\begin{equation}
F_{\mu\nu}=\nabla_\mu A_\nu-\nabla_\nu A_\mu,
\qquad
D_\mu\Psi=\nabla_\mu\Psi-iqA_\mu\Psi .
\label{2.2}
\end{equation}
This is the Abelian-Higgs sector used in the original gravitational mechanism for black hole superconductivity and in the minimal holographic-superconductor construction \cite{Gubser:2008px,Hartnoll:2008vx,Hartnoll:2008kx}.  Varying the matter fields gives
\begin{equation}
D_\mu D^\mu\Psi-m^2\Psi=0,
\qquad
\nabla_\mu F^{\mu\nu}
=iq\left[\Psi^\ast D^\nu\Psi-\Psi(D^\nu\Psi)^\ast\right] .
\label{2.3}
\end{equation}
The scalar mass is allowed to be negative, provided the asymptotic AdS$_4$ vacuum remains stable.  The ultraviolet stability condition is the Breitenlohner-Freedman bound
\begin{equation}
m^2L^2\geq-\frac{9}{4},
\qquad
\Delta_\pm=\frac{3}{2}\pm\frac{1}{2}\sqrt{9+4m^2L^2}.
\label{2.4}
\end{equation}
The two roots $\Delta_\pm$ determine the possible near-boundary falloffs of the scalar.  When both quantizations are allowed, choosing one or the other corresponds to choosing the boundary theory in which the charged operator has dimension $\Delta_+$ or $\Delta_-$ \cite{Breitenlohner:1982bm,Klebanov:1999tb}.  In the present work we use the standard dimension-two quantization associated with $m^2L^2=-2$, because this is the choice used in the canonical holographic Josephson construction \cite{Horowitz:2011dz}.

The action in Eq. \eqref{2.1} admits two related approximations.  In the strict probe limit, one rescales $A_\mu$ and $\Psi$ by inverse powers of $q$ and takes $q\rightarrow\infty$ while keeping the rescaled fields fixed; then the matter fields propagate on a prescribed black hole geometry.  In the charged-background treatment, one keeps the RN-AdS charge sector explicit and treats the scalar condensate as the order parameter that may destabilize the charged normal phase.  These two descriptions overlap when the scalar and gauge perturbations responsible for superconductivity are small enough that their stress tensor does not appreciably change the metric.  A convenient way to state this requirement is
\begin{equation}
\kappa_4^2 L^2
\max\left(
|F_{\mu\nu}F^{\mu\nu}|,
|D_\mu\Psi D^\mu\Psi|,
|m^2||\Psi|^2
\right)\ll 1 .
\label{2.5}
\end{equation}
This condition will be assumed whenever we discuss the RN-AdS geometry as fixed background observable.  It also marks the regime in which very low-temperature or strongly condensed solutions should be interpreted cautiously, since the condensate can eventually carry enough energy to require full backreaction.

\subsection{RN-AdS black brane and scalar instability}
The normal finite-density state is represented by the planar Reissner--Nordstrom-AdS black brane with vanishing scalar field.  We use coordinates in which
\begin{equation}
ds^2=-f(r)dt^2+\frac{dr^2}{f(r)}
+r^2(dx^2+dy^2),
\qquad
A=A_t(r)\,dt,
\qquad
\Psi=0 .
\label{2.6}
\end{equation}
With the normalization of Eq. \eqref{2.1}, a convenient parametrization of the charged solution is \cite{Faulkner:2009wj}
\begin{equation}
f(r)=\frac{r^2}{L^2}-\frac{2M}{r}+\frac{Q^2}{4r^2},
\qquad
A_t(r)=Q\left(\frac{1}{r_h}-\frac{1}{r}\right),
\qquad
\mu=\frac{Q}{r_h}.
\label{2.7}
\end{equation}
The horizon radius $r_h$ is the largest positive root of $f(r_h)=0$.  Imposing this condition fixes the mass parameter as
\begin{equation}
2M=\frac{r_h^3}{L^2}+\frac{Q^2}{4r_h}.
\label{2.8}
\end{equation}
The Hawking temperature follows from the regularity of the Euclidean section, equivalently from $T=f'(r_h)/(4\pi)$, and is
\begin{equation}
T=\frac{1}{4\pi}
\left(
\frac{3r_h}{L^2}
-\frac{Q^2}{4r_h^3}
\right).
\label{2.9}
\end{equation}
Extremality is reached when $T=0$, or
\begin{equation}
Q_{\rm ext}^2=\frac{12r_h^4}{L^2}.
\label{2.10}
\end{equation}
At extremality, the near-horizon expansion gives
\begin{equation}
f(r)=\frac{(r-r_h)^2}{L_2^2}+O\!\left((r-r_h)^3\right),
\qquad
L_2^2=\frac{L^2}{6},
\qquad
A_t(r)=E_h(r-r_h)+O\!\left((r-r_h)^2\right),
\label{2.11}
\end{equation}
where $E_h=Q/r_h^2$.  Thus the infrared region of the extremal planar RN-AdS solution is AdS$_2\times\mathbb{R}^2$, with AdS$_2$ radius $L_2=L/\sqrt{6}$.  This throat is the geometric origin of the low-energy scaling effects that will later be connected to Josephson transport \cite{Faulkner:2009wj}.

The charged scalar instability can be seen already at the linearized level.  Let $\Psi=\psi(r)$ be real, static, and infinitesimal.  Then Eq. \eqref{2.3} reduces to
\begin{equation}
\frac{1}{r^2}\partial_r\!\left(r^2 f\,\partial_r\psi\right)
-\left(
m^2-\frac{q^2A_t^2}{f}
\right)\psi=0 .
\label{2.12}
\end{equation}
This equation shows that the gauge potential lowers the effective radial mass of the scalar.  Equivalently,
\begin{equation}
m_{\rm eff}^2(r)=m^2+q^2g^{tt}A_t^2
=m^2-\frac{q^2A_t^2}{f(r)} .
\label{2.13}
\end{equation}
Since $g^{tt}<0$ outside the horizon, the electric potential contributes negatively to $m_{\rm eff}^2$.  The instability is therefore not caused by violating the AdS$_4$ ultraviolet BF bound in Eq. \eqref{2.4}; rather, it is an infrared instability driven by the finite-density black hole background \cite{Gubser:2008px}.  In the extremal limit, the infrared BF criterion is
\begin{equation}
m_{\rm IR}^2L_2^2
=
\left(m^2-q^2E_h^2L_2^2\right)L_2^2
<-\frac{1}{4}.
\label{2.14}
\end{equation}
For the extremal planar solution in the normalization of Eq. \eqref{2.7}, this becomes $m_{\rm IR}^2L_2^2=m^2L^2/6-q^2L^2/3$.  The important point is that a scalar can be stable with respect to the AdS$_4$ boundary while becoming unstable in the near-horizon AdS$_2$ region.  The onset of condensation is then detected by a normalizable, source-free scalar mode satisfying regularity at the horizon and the chosen AdS boundary condition.

The gauge choice in Eq. \eqref{2.7} is not merely conventional.  The one-form $A=A_tdt$ must be regular at the future horizon, and this requires $A_t(r_h)=0$ in the static gauge used above.  If $A_t$ were nonzero at the horizon, a nonzero charged scalar would require a compensating time-dependent phase, and that phase would be ill-defined on the horizon.  Horizon regularity is therefore part of the definition of the superconducting bulk solution, not an auxiliary assumption.

\subsection{Boundary quantization and superconducting order}
Near the AdS$_4$ boundary, the scalar has the expansion
\begin{equation}
\Psi(r,x)
=
\frac{\psi_-(x)}{r^{\Delta_-}}
+
\frac{\psi_+(x)}{r^{\Delta_+}}
+\cdots ,
\qquad
\Delta_\pm=\frac{3}{2}\pm\frac{1}{2}\sqrt{9+4m^2L^2}.
\label{2.15}
\end{equation}
For $m^2L^2=-2$, one has $\Delta_-=1$ and $\Delta_+=2$, so that
\begin{equation}
\Psi(r,x)=\frac{\psi^{(1)}(x)}{r}
+\frac{\psi^{(2)}(x)}{r^2}
+\cdots .
\label{2.16}
\end{equation}
We impose the source-free dimension-two boundary condition
\begin{equation}
\psi^{(1)}(x)=0,
\qquad
\langle\mathcal{O}_2(x)\rangle=\psi^{(2)}(x),
\label{2.17}
\end{equation}
where the overall normalization of $\mathcal{O}_2$ has been absorbed into the definition of the operator.  This convention matches the standard holographic Josephson setup and allows the condensate profile to be compared directly with the weak-link observable \cite{Horowitz:2011dz}.  A nonzero $\langle\mathcal{O}_2\rangle$ in the absence of the source $\psi^{(1)}$ is the boundary signal of spontaneous breaking of the global $U(1)$ symmetry associated with the bulk Maxwell field.

The Maxwell field determines the finite-density and transport observable of the boundary theory.  In the static configurations relevant to the DC Josephson problem, the asymptotic form is
\begin{equation}
A_t(r,x)=\mu(x)-\frac{\rho(x)}{r}+\cdots,
\qquad
M_i(r,x)=\nu_i(x)+\frac{J_i(x)}{r}+\cdots ,
\label{2.18}
\end{equation}
where, consistently with the charged phase convention used below,
\begin{equation}
M_\mu=A_\mu-\frac{1}{q}\partial_\mu\varphi
\end{equation}
is the gauge-invariant combination obtained after writing the scalar as
\(\Psi=|\Psi|e^{i\varphi}\).  The boundary dictionary is then
\begin{equation}
\mu=A_t^{(0)},
\qquad
\rho=-A_t^{(1)},
\qquad
\nu_i=M_i^{(0)},
\qquad
J_i=M_i^{(1)} .
\label{2.19}
\end{equation}
Here $\mu$ is the chemical potential, $\rho$ is the charge density, $\nu_i$ is the superfluid velocity, and $J_i$ is the conserved boundary current.  The radial Maxwell constraint implies the boundary conservation law
\begin{equation}
\partial_iJ^i=0 .
\label{2.20}
\end{equation}
For the one-dimensional junction considered later, this means that the current across the weak link is independent of the boundary coordinate $x$.  This fact is crucial: it permits us to prescribe a stationary current and then compute the gauge-invariant phase difference across the junction.

The superconducting transition is identified by the appearance of a regular, normalizable, source-free solution for the charged scalar.  In a homogeneous probe-limit superconductor with the dimension-two quantization, the critical temperature is proportional to the chemical potential, and in the normalization used in the original Josephson construction one has $T_c\simeq0.0588\,\mu$ \cite{Horowitz:2011dz}.  In the RN-AdS formulation, the precise critical surface depends on the charge sector and on the scalar parameters, but the physical criterion is the same: a region is superconducting when the source-free condensate is nonzero and normal when the corresponding scalar mode is absent.  This local superconducting criterion will be used to engineer a weak link by choosing an inhomogeneous boundary chemical potential.

\section{Weak-link construction on a charged AdS background}
\subsection{Inhomogeneous chemical potential and SNS regime}
We now deform the homogeneous charged superconductor by imposing a static spatial profile for the boundary chemical potential.  The junction is taken to lie along the $x$ direction, while the $y$ direction remains translationally invariant.  The profile is chosen so that the asymptotic regions $x\rightarrow\pm\infty$ are superconducting at the same temperature, whereas the central region has a reduced local chemical potential and can therefore be driven into the normal phase.  A convenient smooth profile is \cite{Horowitz:2011dz}
\begin{equation}
\mu(x)=\mu_\infty
\left[
1-\frac{1-\epsilon}{2\tanh\!\left(\ell/2\sigma\right)}
\left\{
\tanh\!\left(\frac{x+\ell/2}{\sigma}\right)
-\tanh\!\left(\frac{x-\ell/2}{\sigma}\right)
\right\}
\right] .
\label{3.1}
\end{equation}
Here $\mu_\infty$ is the chemical potential in the two exterior superconducting regions, $\ell$ is the nominal width of the weak link, $\sigma$ controls the smoothness of the interpolation, and $0<\epsilon<1$ controls the depth of the central suppression.  The normalization in Eq. \eqref{3.1} gives
\(\mu(\pm\infty)=\mu_\infty\) and
\(\mu(0)=\epsilon\mu_\infty\) exactly.  For
\(\ell\gg\sigma\), the central plateau remains close to this suppressed value over most of the weak-link region.  Thus the profile creates a finite-density analogue of an SNS junction rather than an externally voltage-biased device.

In the probe-limit dimension-two superconductor used as the canonical weak-link model, the homogeneous critical temperature is proportional to the chemical potential.  We shall write this local estimate as
\begin{equation}
T_c(x)=\alpha\,\mu(x),
\qquad
T_c^\infty=\alpha\,\mu_\infty,
\qquad
T_0=\alpha\,\mu(0)=\epsilon T_c^\infty,
\label{3.2}
\end{equation}
where $\alpha\simeq0.0588$ in the normalization of the standard holographic Josephson junction.  The same notation remains useful in the RN-AdS extension, even when \(\alpha\) is replaced by a charge-sector-dependent critical surface, because the weak-link interpretation depends only on the relative ordering of the corresponding local critical scales.  In that more general case, Eq. \eqref{3.2} should be understood as a design criterion for the boundary source rather than as an independent local-equilibrium phase transition.  The clean SNS regime is
\begin{equation}
T_0<T<T_c^\infty .
\label{3.3}
\end{equation}
For $T<T_0$, the middle region is not normal but weakly superconducting, and the system is better interpreted as an S-S$'$-S junction.  For $T>T_c^\infty$, the condensate disappears in the exterior regions as well, and there is no Josephson weak link.  Hence Eq. \eqref{3.3} is the regime in which a sinusoidal current-phase relation and an exponentially suppressed critical current are expected to be most sharply defined.

\begin{figure}
    \centering
    \includegraphics[width=0.6\textwidth]{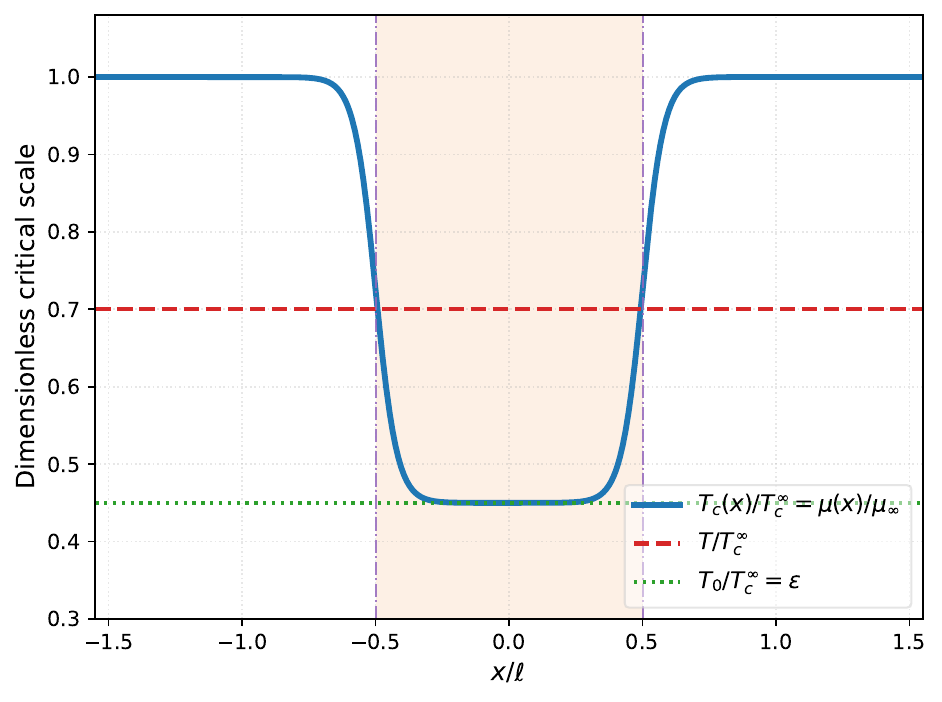}
    \caption{Weak-link chemical-potential profile and SNS operating regime. The blue curve shows the dimensionless local critical scale $T_c(x)/T_c^\infty=\mu(x)/\mu_\infty$ generated by the smooth boundary profile in Eq. \eqref{3.1}, with representative parameters $\epsilon=0.45$ and $\sigma/\ell=0.08$. The shaded interval marks the nominal weak-link region $|x|<\ell/2$. The horizontal dashed line indicates the operating temperature $T/T_c^\infty=0.70$, while the dotted line gives the central scale $T_0/T_c^\infty=\epsilon$. Since $T_0<T<T_c^\infty$, the exterior regions are superconducting whereas the central region is driven normal, realizing the SNS regime described by Eq. \eqref{3.3}.}
    \label{fig_weak-link-profile}
\end{figure}

Figure \ref{fig_weak-link-profile} illustrates how the boundary source realizes the SNS hierarchy. Because the local critical scale is proportional to the imposed chemical potential, the profile $T_c(x)/T_c^\infty=\mu(x)/\mu_\infty$ remains equal to unity in the two exterior banks and is smoothly suppressed to $T_0/T_c^\infty=\epsilon$ at the center of the weak link. For the representative operating temperature shown in the figure, the inequality $T_0<T<T_c^\infty$ is satisfied: the superconducting banks remain below their local critical temperature, while the middle region lies above its suppressed central critical scale. The shaded interval therefore identifies the nominal normal barrier through which the condensate must leak by proximity. This makes explicit that the weak link is produced by an inhomogeneous boundary chemical-potential profile, while the Josephson phase difference and current must still be extracted from the gauge-invariant condensate and Maxwell observable of the full bulk solution.

In the present construction, the phrase \textit{charged AdS background} should be understood in this controlled sense.  The homogeneous RN-AdS charge sector fixes the finite-density normal-state geometry, while the spatially dependent weak-link profile is imposed through the boundary value of the Maxwell-sector variables that determine the superconducting response.  Within the fixed-background treatment, the inhomogeneous fields are therefore treated as probe observable, or equivalently as perturbations whose stress tensor is small in the sense of Eq. \eqref{2.5}.  A fully backreacted inhomogeneous solution would replace the planar RN-AdS metric by a metric depending on both \(r\) and \(x\), but that extension is outside the approximation used here.

The phrase \textit{local critical temperature} in Eq. \eqref{3.2} should not be read as a local-equilibrium approximation to a thermodynamic phase transition at each point.  It is instead a controlled way of designing boundary sources: the chemical potential is chosen so that the exterior region supports a source-free condensate while the central region suppresses it.  The actual condensate profile is determined by the coupled bulk equations and automatically includes proximity leakage into the barrier.  This distinction is essential because the Josephson current is carried precisely by the nonlocal overlap of the superconducting order through the weak link.

\subsection{Gauge-invariant phase variables}
To describe a stationary Josephson state, the scalar phase must be retained.  We write
\begin{equation}
\Psi(r,x)=\psi(r,x)e^{i\varphi(r,x)},
\qquad
A=A_t(r,x)\,dt+A_r(r,x)\,dr+A_x(r,x)\,dx ,
\label{3.4}
\end{equation}
with real functions $\psi$, $\varphi$, $A_t$, $A_r$, and $A_x$.  Since the scalar charge $q$ is kept explicit, the gauge-invariant vector field is
\begin{equation}
M_\mu=A_\mu-\frac{1}{q}\partial_\mu\varphi,
\qquad
D_\mu\Psi=e^{i\varphi}\left(\partial_\mu\psi-iqM_\mu\psi\right).
\label{3.5}
\end{equation}
Under a bulk gauge transformation $A_\mu\rightarrow A_\mu+\partial_\mu\lambda$ and $\varphi\rightarrow\varphi+q\lambda$, the amplitude $\psi$ and the field $M_\mu$ are invariant.  Therefore all stationary observables of the weak link must be expressible in terms of $\psi$ and $M_\mu$, not in terms of the gauge-dependent phase $\varphi$ alone.

Substituting Eq. \eqref{3.5} into the scalar equation and Maxwell equations in the metric of Eq. \eqref{2.6}, allowing dependence only on $(r,x)$, gives the gauge-invariant stationary system
\begin{equation}
\begin{aligned}
0={}&
\partial_r^2\psi+
\left(\frac{f'}{f}+\frac{2}{r}\right)\partial_r\psi
+\frac{1}{r^2f}\partial_x^2\psi
+\left(
\frac{q^2M_t^2}{f^2}
-q^2M_r^2
-\frac{q^2M_x^2}{r^2f}
-\frac{m^2}{f}
\right)\psi,\\
0={}&
\partial_r^2M_t+\frac{2}{r}\partial_rM_t
+\frac{1}{r^2f}\partial_x^2M_t
-\frac{2q^2\psi^2}{f}M_t,\\
0={}&
\partial_x^2M_r-\partial_r\partial_xM_x
-2q^2r^2\psi^2M_r,\\
0={}&
\partial_r^2M_x-\partial_r\partial_xM_r
+\frac{f'}{f}\left(\partial_rM_x-\partial_xM_r\right)
-\frac{2q^2\psi^2}{f}M_x,\\
0={}&
\partial_rM_r+\frac{1}{r^2f}\partial_xM_x
+\frac{2}{\psi}\left(
M_r\partial_r\psi+\frac{M_x}{r^2f}\partial_x\psi
\right)
+\left(\frac{f'}{f}+\frac{2}{r}\right)M_r .
\end{aligned}
\label{3.6}
\end{equation}
The first line is the real part of the charged scalar equation.  The next three lines are the $t$, $r$, and $x$ components of Maxwell's equation.  The last line is the phase equation, equivalently the conservation law $\nabla_\mu(\psi^2M^\mu)=0$.  The system is elliptic for stationary configurations outside the horizon, supplemented by one first-order constraint.  No separate equation for $\varphi$ appears because its only invariant content has been absorbed into $M_\mu$.

The construction is symmetric under $x\rightarrow -x$ when the chemical-potential profile is even and the two exterior superconductors are identical.  For the branch relevant to a uniform current through the junction, one may choose parity assignments
\begin{equation}
\psi,\;M_t,\;M_x\ \text{even in }x,
\qquad
M_r\ \text{odd in }x .
\label{3.7}
\end{equation}
These assignments are not additional physics; they are a convenient way of selecting the symmetric SNS configuration.  They ensure that the two superconducting banks are identical and that the phase bias is entirely encoded in the gauge-invariant spatial component $M_x$.

\subsection{Boundary observable and phase difference}
At the AdS boundary the fields admit the asymptotic expansions
\begin{equation}
\psi(r,x)=\frac{\psi^{(1)}(x)}{r}
+\frac{\psi^{(2)}(x)}{r^2}
+\cdots,
\qquad
M_t(r,x)=\mu(x)-\frac{\rho(x)}{r}+\cdots,
\qquad
M_x(r,x)=\nu(x)+\frac{J(x)}{r}+\cdots .
\label{3.8}
\end{equation}
We impose the source-free scalar boundary condition $\psi^{(1)}(x)=0$ and identify $\psi^{(2)}(x)$ with the superconducting condensate.  The prescribed chemical potential is the weak-link profile in Eq. \eqref{3.1}.  The leading spatial component $\nu(x)$ is the boundary superfluid velocity, while the coefficient $J(x)$ is the current density across the junction.

The radial component has no independent boundary source in the stationary weak-link problem and decays sufficiently fast near the boundary.  We impose
\begin{equation}
M_r(r,x)=\mathcal{O}(r^{-3})
\qquad
(r\rightarrow\infty).
\label{3.9}
\end{equation}
This condition is compatible with the asymptotic Maxwell constraint and removes an otherwise spurious radial source.  Expanding the last line of Eq. \eqref{3.6} near the boundary gives
\begin{equation}
\partial_xJ(x)=0 .
\label{3.10}
\end{equation}
Thus the current through a stationary one-dimensional junction is constant.  We shall therefore write $J(x)=J$ and use $J$ as the control parameter for the DC Josephson branch.

At the horizon, regularity of the one-form $M_tdt$ requires
\begin{equation}
M_t(r_h,x)=0 .
\label{3.11}
\end{equation}
The remaining fields are required to be smooth in coordinates regular at the future horizon.  These regularity conditions determine the admissible radial derivatives of $\psi$, $M_r$, and $M_x$ once the horizon values are specified consistently with Eq. \eqref{3.6}.  This is the charged-background version of the standard holographic-superconductor horizon condition: a nonzero temporal gauge potential at the horizon would force an ill-defined time-dependent scalar phase.

In the boundary spatial direction, the fields approach homogeneous superconductors on both sides of the junction:
\begin{equation}
\lim_{x\rightarrow\pm\infty}
\left\{
\psi(r,x),M_t(r,x),M_r(r,x),M_x(r,x)
\right\}
=
\left\{
\psi_{\rm hom}(r),M_{t,{\rm hom}}(r),0,M_{x,{\rm hom}}(r)
\right\}.
\label{3.12}
\end{equation}
The equality of the two asymptotic superconductors means that $\mu(+\infty)=\mu(-\infty)=\mu_\infty$ and that the imposed current is the same on both sides.  A nonzero current generally induces a nonzero asymptotic superfluid velocity, so the phase difference must be defined with this homogeneous contribution subtracted.

The gauge-invariant phase difference across the junction is \cite{Horowitz:2011dz}
\begin{equation}
\gamma
=
\varphi(+\infty)-\varphi(-\infty)
-q\int_{-\infty}^{+\infty} A_x(\infty,x)\,dx .
\label{3.13}
\end{equation}
Using Eq. \eqref{3.5}, this becomes
\begin{equation}
\gamma
=
-q\int_{-\infty}^{+\infty}
\left[\nu(x)-\nu(\infty)\right]dx ,
\label{3.14}
\end{equation}
where $\nu(+\infty)=\nu(-\infty)\equiv\nu(\infty)$ for the symmetric current-carrying branch.  The subtraction makes the integral finite by removing the phase gradient of the homogeneous superconducting banks.  In practice, one fixes $J$, solves Eq. \eqref{3.6} with the boundary conditions above, extracts $\nu(x)$ from Eq. \eqref{3.8}, and then computes $\gamma$ from Eq. \eqref{3.14}.  The Josephson relation is then obtained parametrically as $J(\gamma)$.

The spatially varying chemical potential should be understood as a static material profile rather than an imposed voltage difference between the two superconductors.  Since $\mu(x)$ changes near the edges of the weak link, the boundary configuration contains a static electric-field profile in the formal sense of the boundary source.  The stationary solution remains meaningful because the bulk equations determine a charge-density profile and a conserved current that balance this inhomogeneity.  The AC Josephson problem, in which a genuine voltage bias produces time dependence of $\gamma$, is a different problem and is not part of the present construction.

\section{Josephson observables in the RN-AdS weak link}
\subsection{Current-phase relation}
The weak-link solution constructed in Sec.~3 defines a one-parameter family of stationary states once the geometric and boundary parameters are fixed.  It is useful to package the RN-AdS charge dependence by introducing the dimensionless charge variable
\begin{equation}
\chi=\frac{Q}{Q_{\rm ext}},
\qquad
0\leq \chi\leq 1,
\qquad
T=\frac{3r_h}{4\pi L^2}\left(1-\chi^2\right),
\label{4.1}
\end{equation}
where $Q_{\rm ext}$ was defined in Eq. \eqref{2.10}.  At fixed $(\chi,T,\ell,\epsilon,\sigma)$, the stationary branch is obtained by prescribing the conserved boundary current $J$ and extracting the gauge-invariant phase difference $\gamma$ from Eq. \eqref{3.14}.  The resulting observable is therefore the parametric curve
\begin{equation}
J\longmapsto \gamma(J;\chi,T,\ell,\epsilon,\sigma),
\qquad
J(\gamma;\chi,T,\ell,\epsilon,\sigma)
=\gamma^{-1}(J)
\quad \text{on a monotonic branch}.
\label{4.2}
\end{equation}
The inverse notation is only local: a current-carrying weak link can possess multiple stationary branches, and the Josephson branch relevant for the DC effect is the one continuously connected to $J=0$ and $\gamma=0$.

Gauge invariance restricts the current-phase relation to be a function of the phase difference modulo $2\pi$.  Time-reversal symmetry at zero imposed voltage implies that reversing the phase bias reverses the current.  Hence the stationary current admits the harmonic decomposition
\begin{equation}
J(\gamma)=\sum_{n=1}^{\infty}J_n\sin(n\gamma),
\qquad
J(-\gamma)=-J(\gamma),
\qquad
J(\gamma+2\pi)=J(\gamma).
\label{4.3}
\end{equation}
In an opaque SNS junction, higher harmonics are suppressed and the leading Josephson relation becomes
\begin{equation}
J(\gamma)=J_c\sin\gamma+\mathcal{O}(\sin2\gamma).
\label{4.4}
\end{equation}
The holographic observable is not the sine law itself, which is already known in the Schwarzschild-AdS weak-link construction \cite{Horowitz:2011dz}, but the dependence of the coefficients $J_n$ on the RN-AdS charge sector.  In particular, the leading diagnostic is
\begin{equation}
J_c(\chi,T,\ell,\epsilon,\sigma)
=\max_{\gamma\in[-\pi,\pi]}|J(\gamma;\chi,T,\ell,\epsilon,\sigma)|.
\label{4.5}
\end{equation}
For a purely sinusoidal branch, $J_c=J_1$.  If near-extremal throat effects or an S-S$'$-S crossover generate appreciable higher harmonics, Eq. \eqref{4.5} remains the operational definition of the critical current, while Eq. \eqref{4.3} provides the sharper decomposition of the deviation from the elementary Josephson form.

The first consistency requirement is that the current-phase relation reduce to the known holographic SNS result when charged-background effects are removed or made parametrically weak.  In the notation above, this means that for a Schwarzschild-AdS-like weak link one must recover a dominant first harmonic and the standard exponential suppression of $J_c$ with the junction width.  The RN-AdS analysis is therefore organized around the deviation
\begin{equation}
\delta J(\gamma;\chi)
=
J(\gamma;\chi)-J(\gamma;0),
\label{4.6}
\end{equation}
where the second term denotes the corresponding uncharged or Schwarzschild-AdS reference branch at matched boundary temperature and weak-link profile.  This subtraction is conceptual rather than a new boundary condition: it isolates the part of the Josephson response controlled by finite-density RN-AdS observable.

\subsection{Critical current and coherence length}
The critical current measures the transparency of the weak link to the superconducting order.  In an SNS junction with a sufficiently opaque normal barrier, the condensate leaking from the two superconducting banks overlaps weakly inside the central region.  The standard expectation is then an exponential suppression of the maximum current with the junction width \cite{Likharev:1979zz,Horowitz:2011dz},
\begin{equation}
J_c(\chi,T,\ell)
=A_J(\chi,T)\exp\!\left[-\frac{\ell}{\xi_J(\chi,T)}\right]
\left[1+o(1)\right],
\qquad
\ell\gg \xi_J .
\label{4.7}
\end{equation}
Here $\xi_J$ is the Josephson coherence length extracted from the decay of the critical current.  In the holographic construction, both $A_J$ and $\xi_J$ are emergent functions of the bulk scalar, Maxwell field, and RN-AdS charge sector.

The same coherence scale can be diagnosed from the condensate at the center of the barrier at zero current.  Since the order parameter must leak from each superconducting bank to the midpoint, the midpoint condensate is expected to decay with half the exponent appearing in Eq. \eqref{4.7}:
\begin{equation}
\langle\mathcal{O}_2\rangle_{x=0,J=0}
=A_{\mathcal O}(\chi,T)
\exp\!\left[-\frac{\ell}{2\xi_{\mathcal O}(\chi,T)}\right]
\left[1+o(1)\right].
\label{4.8}
\end{equation}
A controlled SNS regime should satisfy
\begin{equation}
\xi_J(\chi,T)=\xi_{\mathcal O}(\chi,T)
\label{4.9}
\end{equation}
within the accuracy allowed by the smoothness of the chemical-potential profile and by the distance from the ideal opaque-barrier limit.  Equation \eqref{4.9} is a nontrivial internal check because $J_c$ and the midpoint condensate are extracted from different boundary observables.

The origin of the relative factor of two in Eqs. \eqref{4.7} and \eqref{4.8} can be made explicit in a one-dimensional effective barrier model.  In the normal region, linearize the order-parameter functional as \cite{Likharev:1979zz}
\begin{equation}
\mathcal F_N
=
\int_{-\ell/2}^{\ell/2}dx
\left[
K|\partial_x\Phi|^2+a|\Phi|^2
\right],
\qquad
a>0,
\qquad
\xi_N=\sqrt{\frac{K}{a}} .
\label{4.10}
\end{equation}
The Euler-Lagrange equation is
\begin{equation}
-\xi_N^2\Phi''+\Phi=0,
\qquad
\Phi(-\ell/2)=\Phi_0e^{-i\gamma/2},
\qquad
\Phi(\ell/2)=\Phi_0e^{i\gamma/2}.
\label{4.11}
\end{equation}
Solving the boundary-value problem gives
\begin{equation}
\Phi(x)=
\Phi_0\frac{\cos(\gamma/2)}{\cosh(\ell/2\xi_N)}
\cosh\!\left(\frac{x}{\xi_N}\right)
+i\Phi_0\frac{\sin(\gamma/2)}{\sinh(\ell/2\xi_N)}
\sinh\!\left(\frac{x}{\xi_N}\right).
\label{4.12}
\end{equation}
The conserved current associated with the phase rotation of $\Phi$ is
\begin{equation}
J_N
=
2qK\,{\rm Im}\!\left(\Phi^\ast\partial_x\Phi\right)
=
\frac{2qK\Phi_0^2}{\xi_N\sinh(\ell/\xi_N)}\sin\gamma .
\label{4.13}
\end{equation}
For $\ell\gg\xi_N$, Eq. \eqref{4.13} becomes $J_N\propto e^{-\ell/\xi_N}\sin\gamma$, while Eq. \eqref{4.12} gives $|\Phi(0)|\propto e^{-\ell/2\xi_N}$ at $J=0$.  The holographic calculation replaces the effective constants $(K,a,\Phi_0)$ by bulk radial dynamics, but the exponential structure and the factor-of-two relation between $J_c$ and the midpoint condensate remain the reference for an SNS interpretation.

\begin{figure}
    \centering
    \includegraphics[width=0.48\textwidth]{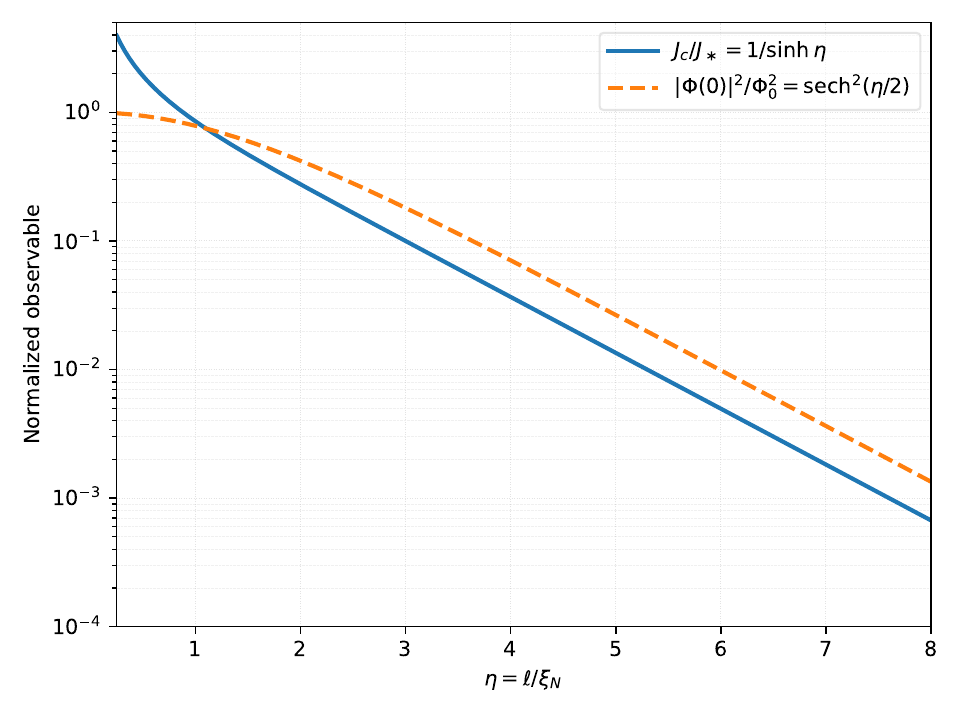}
    \includegraphics[width=0.48\textwidth]{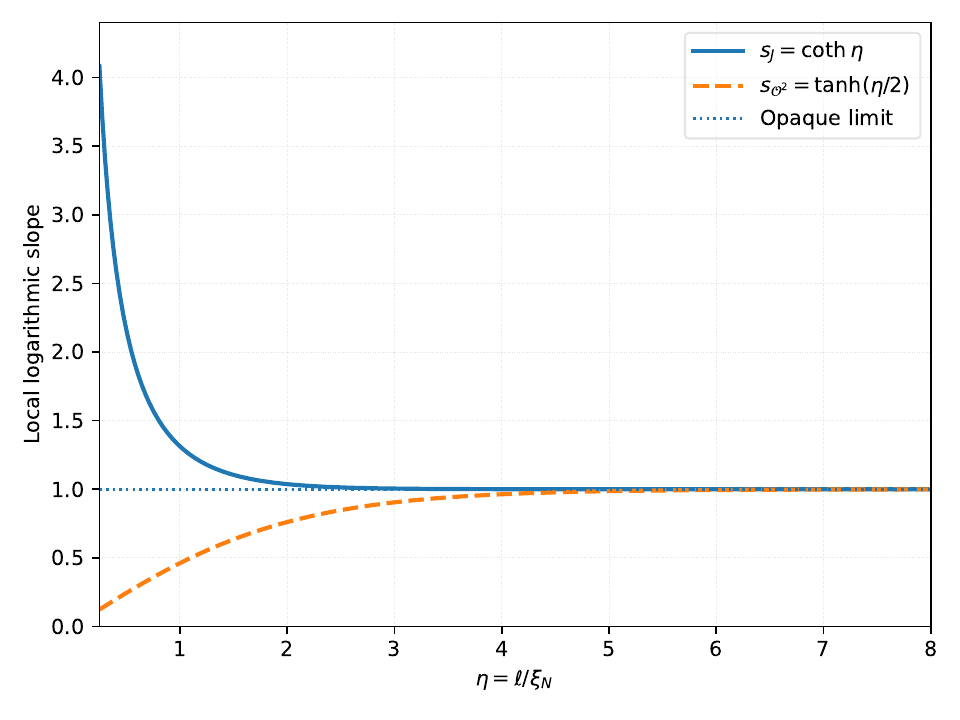}
    \caption{Exact proximity reference for coherence-length extraction in the linear normal-barrier model. The left panel compares the critical-current scale $J_c/J_\ast=1/\sinh(\ell/\xi_N)$ with the squared midpoint condensate $|\Phi(0)|^2/\Phi_0^2=\sech^2(\ell/2\xi_N)$. The right panel shows the corresponding local logarithmic slopes, $s_J=\coth(\ell/\xi_N)$ and $s_{\mathcal O^2}=\tanh(\ell/2\xi_N)$, which approach the common opaque-barrier value as $\ell/\xi_N\rightarrow\infty$.}
    \label{fig_exact_proximity_reference}
\end{figure}

Figure \ref{fig_exact_proximity_reference} isolates the finite-width structure behind the coherence-length criterion used for the holographic SNS junction. In the exact linear barrier solution, the critical current is controlled by $1/\sinh(\ell/\xi_N)$, whereas the squared midpoint condensate is controlled by $\sech^2(\ell/2\xi_N)$. These quantities are not identical at finite width, but their logarithmic slopes converge to the same value in the opaque-barrier regime. The right panel makes this convergence explicit: $s_J$ approaches unity from above, while $s_{\mathcal O^2}$ approaches unity from below. Therefore, agreement between the coherence length extracted from $J_c$ and that extracted from $\langle\mathcal O_2\rangle_{x=0,J=0}^2$ should be interpreted as an asymptotic SNS diagnostic rather than as an exact finite-width identity. In the full RN-AdS weak link, deviations from this reference would indicate finite-transparency corrections, departure from the normal-barrier regime, or additional radial scaling by the charged background.

The RN-AdS-specific question is whether $\xi_J$ and $\xi_{\mathcal O}$ acquire systematic dependence on $\chi$.  If the charged horizon merely changes the local condensate amplitude, then the main effect is a renormalization of $A_J$ and $A_{\mathcal O}$.  If the near-horizon charge sector changes the radial scaling of the scalar mode, then the exponent itself can shift, producing
\begin{equation}
\frac{\partial \xi_J}{\partial \chi}\neq 0,
\qquad
\frac{\partial \xi_{\mathcal O}}{\partial \chi}\neq 0 .
\label{4.14}
\end{equation}
This is the first place where RN-AdS charge can enter as more than a background density: it can modify the effective coherence scale controlling phase-coherent tunneling through the weak link.

\subsection{Phase stiffness and charge-sector response}
The small-phase response defines the Josephson phase stiffness of the weak link.  We define
\begin{equation}
K_J(\chi,T,\ell,\epsilon,\sigma)
=
\left.
\frac{\partial J}{\partial\gamma}
\right|_{\gamma=0}.
\label{4.15}
\end{equation}
For a purely sinusoidal branch, $K_J=J_c$.  More generally, using Eq. \eqref{4.3},
\begin{equation}
K_J=\sum_{n=1}^{\infty}nJ_n .
\label{4.16}
\end{equation}
Thus $K_J$ is sensitive to higher harmonics even when the critical current is dominated by the first harmonic.  It is therefore a useful diagnostic for distinguishing an ordinary opaque SNS regime from a more transparent S-S$'$-S regime or a throat-renormalized near-extremal regime.

The stiffness can also be expressed as the curvature of the phase-dependent part of the stationary free energy.  Let $\Omega(\gamma)$ denote the grand potential of the stationary branch at fixed boundary chemical-potential profile and temperature.  The DC Josephson relation is equivalently
\begin{equation}
J(\gamma)=\frac{\partial\Omega}{\partial\gamma},
\qquad
K_J=
\left.
\frac{\partial^2\Omega}{\partial\gamma^2}
\right|_{\gamma=0},
\label{4.17}
\end{equation}
up to the conventional normalization of the boundary charge.  For the leading Josephson energy $\Omega(\gamma)=\Omega_0+J_c(1-\cos\gamma)$, Eq. \eqref{4.17} reproduces Eq. \eqref{4.4}.  In the fully holographic description, $\Omega$ is obtained from the renormalized on-shell action, but the derivative definition remains independent of the details of the subtraction scheme because only the phase-dependent part is used.

This stiffness is the weak-link analogue of the superfluid density in the homogeneous holographic superconductor.  In a homogeneous system, a static boundary superfluid velocity $\nu_x$ induces a current whose linear response defines a phase rigidity.  In the junction, the imposed phase difference is localized across the weak link, so the corresponding rigidity is nonlocal and depends on the barrier profile.  The comparison may be summarized as
\begin{equation}
J_{\rm hom}=n_s\,\nu_x+\mathcal{O}(\nu_x^3),
\qquad
J_{\rm JJ}=K_J\,\gamma+\mathcal{O}(\gamma^3).
\label{4.18}
\end{equation}
The first relation probes the bulk condensate in a translationally invariant background, while the second probes the ability of the inhomogeneous condensate to transmit phase coherence across the suppressed region.

Near the superconducting critical temperature, the source-free condensate vanishes continuously in the minimal model, and the phase stiffness must vanish with it.  If the transition remains mean-field-like, the condensate amplitude scales as $(1-T/T_c)^{1/2}$, while the phase stiffness and critical current scale quadratically in the order-parameter amplitude.  Thus one expects
\begin{equation}
K_J\rightarrow0,
\qquad
J_c\rightarrow0,
\qquad
T\rightarrow T_c^- .
\label{4.19}
\end{equation}
The precise power can be distorted by the inhomogeneous profile and by the RN-AdS charge sector, but the vanishing itself is required: without a source-free condensate in the superconducting banks, there is no gauge-invariant Josephson phase difference to sustain a nondissipative weak-link current.

Equations \eqref{4.15}--\eqref{4.19} give the observables that will be used to separate three physical effects.  The first is the ordinary proximity suppression due to the width and depth of the weak link.  The second is the finite-density modification of the condensate and superfluid response by the RN-AdS charge.  The third, isolated more sharply in Sec.~5, is the additional infrared scaling caused by the near-extremal AdS$_2\times\mathbb{R}^2$ throat.  Only the third effect is intrinsically tied to near-extremal RN-AdS geometry.

\section{Near-extremal throat control of Josephson transport}
\subsection{AdS \texorpdfstring{$_2\times\mathbb{R}^2$}{} scaling regime}
We now isolate the part of the weak-link response that is specific to near-extremal RN-AdS geometry.  Let
\begin{equation}
\chi=\frac{Q}{Q_{\rm ext}},
\qquad
\varepsilon_T=1-\chi^2
=\frac{4\pi L^2T}{3r_h},
\qquad
\zeta=r-r_h .
\label{5.1}
\end{equation}
The near-extremal regime is $\varepsilon_T\ll1$, or equivalently $T\ll r_h/L^2$ at fixed $r_h$.  Expanding the blackening factor of Eq. \eqref{2.7} near the horizon gives \cite{Faulkner:2009wj}
\begin{equation}
f(r_h+\zeta)
=
4\pi T\,\zeta+\frac{\zeta^2}{L_2^2}
+O\!\left(\frac{\zeta^3}{r_hL^2},\varepsilon_T\frac{\zeta^2}{L^2}\right),
\qquad
L_2^2=\frac{L^2}{6}.
\label{5.2}
\end{equation}
Equivalently, if
\begin{equation}
\zeta_T=4\pi T L_2^2=\frac{r_h}{2}(1-\chi^2),
\label{5.3}
\end{equation}
then the near-horizon metric is, to leading order,
\begin{equation}
ds^2
=
-\frac{\zeta(\zeta+\zeta_T)}{L_2^2}\,dt^2
+\frac{L_2^2\,d\zeta^2}{\zeta(\zeta+\zeta_T)}
+r_h^2(dx^2+dy^2).
\label{5.4}
\end{equation}
This is the finite-temperature AdS$_2$ black hole times $\mathbb{R}^2$.  The scale $\zeta_T$ cuts off the throat in the infrared, while the matching to the asymptotic AdS$_4$ region occurs at some $\zeta_{\rm UV}$ satisfying $\zeta_T\ll \zeta_{\rm UV}\ll r_h$.  The proper radial length of the scaling region is therefore
\begin{equation}
\mathcal L_{\rm throat}
=
L_2\int_{\zeta_T}^{\zeta_{\rm UV}}
\frac{d\zeta}{\sqrt{\zeta(\zeta+\zeta_T)}}
=
2L_2
\left[
\sinh^{-1}\!\sqrt{\frac{\zeta}{\zeta_T}}
\right]_{\zeta_T}^{\zeta_{\rm UV}} .
\label{5.5}
\end{equation}
For \(\zeta_{\rm UV}\gg\zeta_T\), this becomes
\begin{equation}
\mathcal L_{\rm throat}
=
L_2\log\!\left(\frac{\zeta_{\rm UV}}{\zeta_T}\right)+\mathcal{O}(L_2).
\end{equation}
Thus the near-extremal limit is not merely the statement that the charge density is large.  It is the statement that the bulk contains a parametrically long radial region through which the charged scalar and Maxwell sector must propagate before reaching the ultraviolet boundary.

The gauge field in the same region has the regular expansion
\begin{equation}
A_t(r_h+\zeta)
=
E_h\zeta+\mathcal{O}(\zeta^2),
\qquad
E_h=\frac{Q}{r_h^2}.
\label{5.6}
\end{equation}
Combining Eqs. \eqref{5.4} and \eqref{5.6}, one finds that the charged scalar experiences a finite electric contribution to its infrared mass.  For a Fourier component with spatial momentum $k$ along the boundary directions, the effective AdS$_2$ mass is
\begin{equation}
m_{k,{\rm IR}}^2
=
m^2+\frac{k^2}{r_h^2}
-q^2E_h^2L_2^2 .
\label{5.7}
\end{equation}
The corresponding infrared index is
\begin{equation}
\nu_k
=
\sqrt{
\frac{1}{4}
+L_2^2\left(m^2+\frac{k^2}{r_h^2}\right)
-q^2E_h^2L_2^4
},
\qquad
\Delta_k^{\rm IR}=\frac{1}{2}+\nu_k .
\label{5.8}
\end{equation}
The zero-momentum index $\nu_0$ controls the most relevant homogeneous scalar response in the throat.  If $\nu_0$ is real, the throat supports a stable infrared scaling mode.  If $\nu_0$ becomes imaginary, the AdS$_2$ BF bound is violated and the RN-AdS normal phase is unstable to charged scalar condensation, consistently with the mechanism described in Sec.~2.

\begin{figure}
    \centering
    \includegraphics[width=0.6\textwidth]{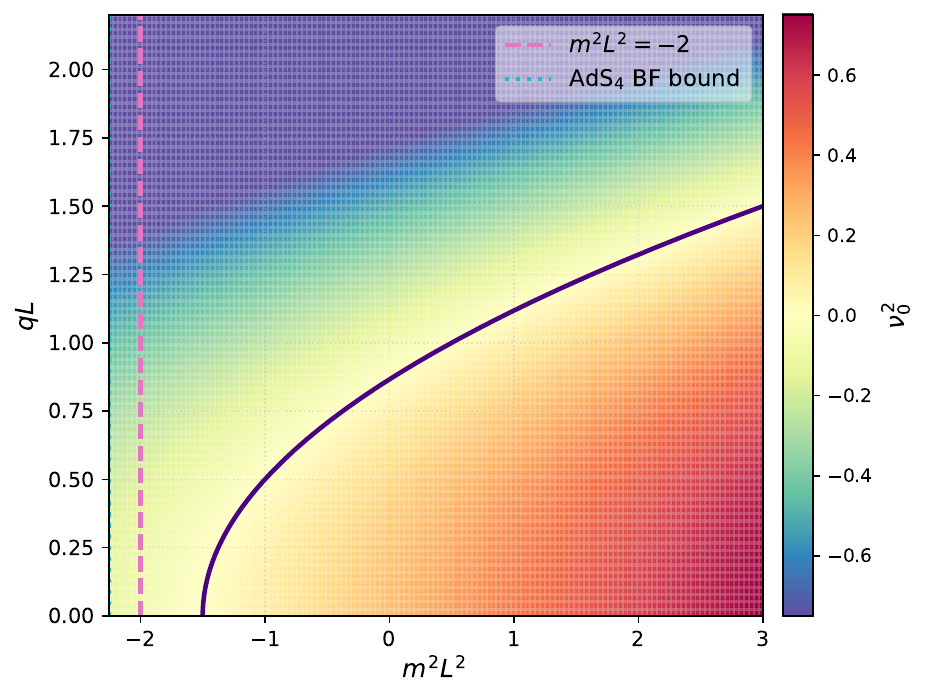}
    \caption{AdS$_2$ infrared index and BF-bound stability map. The color scale shows $\nu_0^2$ for the zero-momentum charged scalar mode in the extremal RN-AdS$_4$ throat, using $\nu_0^2=1/4+m^2L^2/6-q^2L^2/3$. The solid contour denotes $\nu_0^2=0$, which is the AdS$_2$ BF threshold. Below this contour, $\nu_0$ is real and the throat supports a stable infrared scaling mode. Above it, $\nu_0$ is imaginary and the RN-AdS normal throat is unstable to charged-scalar condensation. The dashed vertical line marks the canonical dimension-two mass $m^2L^2=-2$, while the dotted line marks the AdS$_4$ BF bound.}
    \label{fig_ads2-ir-index-map}
\end{figure}

Figure \ref{fig_ads2-ir-index-map} shows that the near-extremal scaling analysis is subject to a nontrivial infrared stability condition. Although the scalar may satisfy the ultraviolet AdS$_4$ BF bound, the electric field of the extremal RN-AdS throat lowers the effective AdS$_2$ mass and can make $\nu_0^2$ negative. The contour $\nu_0^2=0$ therefore separates stable radial transmission through the throat from the regime in which the normal near-horizon geometry is already unstable. This distinction is important for the Josephson problem: the residual scaling ansatz for $\mathcal R_{\rm IR}$ assumes real $\nu_0$, whereas an imaginary index should instead be interpreted as the onset of scalar condensation in the infrared throat. In particular, the canonical choice $m^2L^2=-2$ lies in the imaginary-$\nu_0$ region at extremality, so any application of the real-index scaling law with this mass requires either moving away from the strict extremal normal throat, modifying the scalar parameters, or treating the condensed near-horizon solution rather than the uncondensed RN-AdS throat.

The radial origin of Eq. \eqref{5.8} is transparent.  In the zero-frequency throat region, the scalar equation reduces to
\begin{equation}
\zeta^2\psi_k''+2\zeta\psi_k'
-
m_{k,{\rm IR}}^2L_2^2\psi_k=0,
\label{5.9}
\end{equation}
whose independent power-law solutions are
\begin{equation}
\psi_k(\zeta)
=
c_-\zeta^{-1/2-\nu_k}
+
c_+\zeta^{-1/2+\nu_k}.
\label{5.10}
\end{equation}
The ratio of a throat-scale amplitude to its ultraviolet matching amplitude therefore contains powers of \(\zeta_T/\zeta_{\rm UV}\).  For a stable mode, the leading logarithmic throat transfer factor has the form
\begin{equation}
\mathcal Z_{\rm IR}(k)
\sim
\exp\!\left[
-2\nu_k\frac{\mathcal L_{\rm throat}^{\rm log}}{L_2}
\right]
=
\left(\frac{\zeta_T}{\zeta_{\rm UV}}\right)^{2\nu_k},
\qquad
\mathcal L_{\rm throat}^{\rm log}
\equiv
L_2\log\!\left(\frac{\zeta_{\rm UV}}{\zeta_T}\right).
\label{5.11}
\end{equation}
This expression is the geometric statement behind the expected nonanalytic near-extremal dependence of boundary observables.  The exponent is not set by the spatial width of the Josephson junction; it is set by radial propagation through the AdS$_2$ throat.

\subsection{IR scaling of the Josephson coupling}
The Josephson branch may be described at small phase difference by a phase-dependent contribution to the stationary grand potential,
\begin{equation}
\Omega_J(\gamma)
=
-E_J\cos\gamma
+\mathcal{O}(\cos2\gamma),
\qquad
J(\gamma)=\frac{\partial\Omega_J}{\partial\gamma}.
\label{5.12}
\end{equation}
In the normalization of Eq. \eqref{4.17}, the leading harmonic gives \(J_c=E_J\).  The task is therefore to identify how the RN-AdS throat modifies the effective Josephson energy.  The ordinary weak-link suppression is spatial and was already encoded in Eq. \eqref{4.7}.  The near-extremal throat contributes an additional radial scaling.  We write this separation as
\begin{equation}
J_c(\chi,T,\ell)
=
A_J(\chi,T)
\exp\!\left[-\frac{\ell}{\xi_J(\chi,T)}\right]
\mathcal R_{\rm IR}(\chi,T).
\label{5.13}
\end{equation}
The first two factors describe the boundary weak link.  The last factor isolates the part of the response associated with the near-horizon scaling region.  In a throat-dominated regime with real $\nu_0$, the leading scaling ansatz is
\begin{equation}
\mathcal R_{\rm IR}(\chi,T)
=
C_{\rm IR}(\chi)
\left(
\frac{\zeta_T}{\zeta_{\rm UV}}
\right)^{2\nu_0}
\left[
1+O\!\left(\frac{\zeta_T}{\zeta_{\rm UV}}\right)
\right].
\label{5.14}
\end{equation}
Using Eq. \eqref{5.3}, this may equivalently be written as a low-temperature power law at fixed ultraviolet matching scale,
\begin{equation}
\mathcal R_{\rm IR}(\chi,T)
=
\widetilde C_{\rm IR}(\chi)
\left(
\frac{T}{\Lambda_{\rm IR}}
\right)^{2\nu_0}
\left[
1+O\!\left(\frac{T}{\Lambda_{\rm IR}}\right)
\right],
\qquad
\Lambda_{\rm IR}\sim\frac{\zeta_{\rm UV}}{4\pi L_2^2}.
\label{5.15}
\end{equation}
Equations \eqref{5.13}--\eqref{5.15} should be read as a scaling hypothesis to be tested within the holographic weak-link solution, not as an automatic consequence of finite charge density.  The ordinary SNS exponential in $\ell$ must first be removed before one can claim that the remaining temperature or charge dependence is controlled by the AdS$_2$ throat.

\begin{figure}
    \centering
    \includegraphics[width=0.48\textwidth]{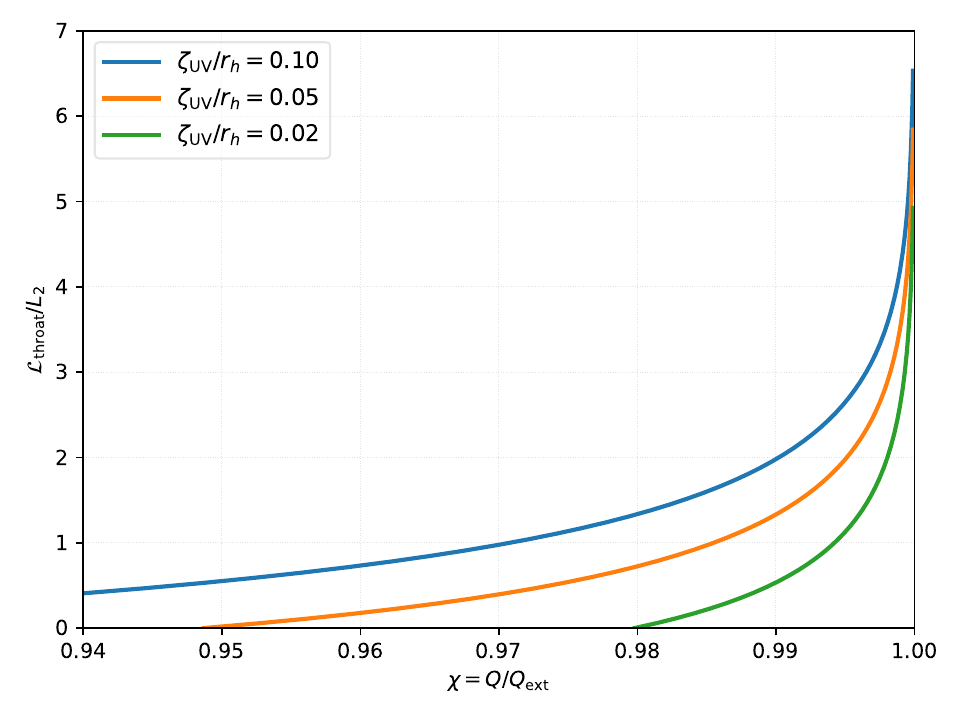}
    \includegraphics[width=0.48\textwidth]{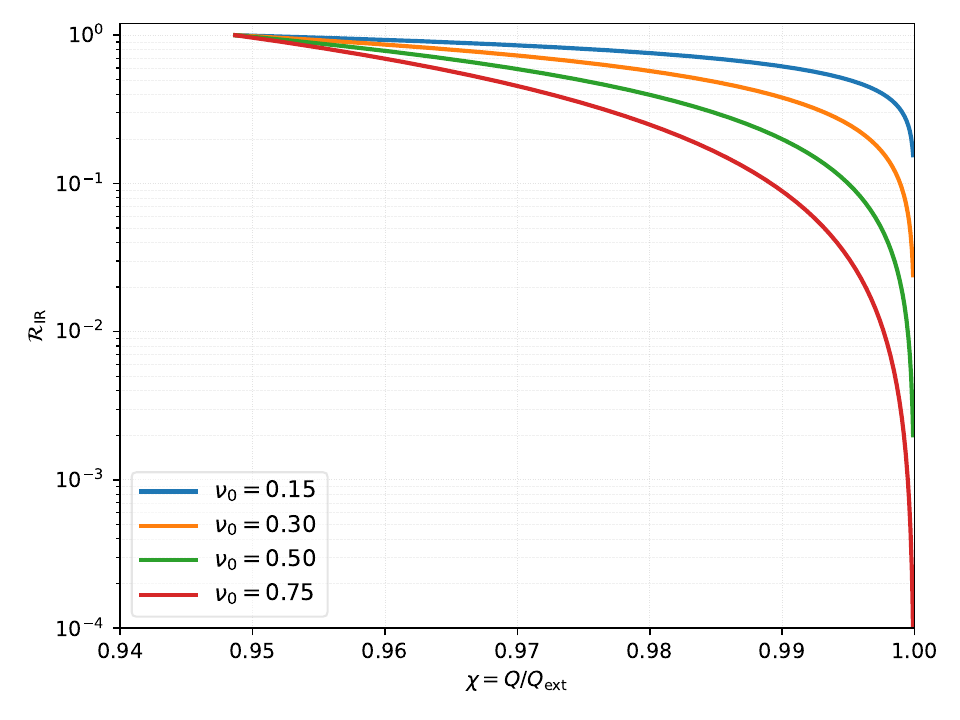}
    \caption{Near-extremal throat growth and residual Josephson factor. The left panel shows the proper AdS$_2$ throat length $\mathcal L_{\rm throat}/L_2$ as a function of the charge ratio $\chi=Q/Q_{\rm ext}$ for representative matching scales $\zeta_{\rm UV}/r_h$. The right panel shows the leading real-index residual factor $\mathcal R_{\rm IR}=(\zeta_T/\zeta_{\rm UV})^{2\nu_0}$ at fixed $\zeta_{\rm UV}/r_h=0.05$ for several stable infrared indices $\nu_0$. Curves are shown only in the hierarchy $\zeta_T<\zeta_{\rm UV}$.}
    \label{fig_throat_residual_scaling}
\end{figure}

Figure \ref{fig_throat_residual_scaling} displays the geometric and transport aspects of the same near-extremal scaling interval. As $\chi$ approaches unity, the infrared cutoff $\zeta_T/r_h=(1-\chi^2)/2$ collapses, opening a radial region between $\zeta_T$ and the matching scale $\zeta_{\rm UV}$ whose proper length grows logarithmically. The residual Josephson factor decreases as a real-index power of $\zeta_T/\zeta_{\rm UV}$, with larger $\nu_0$ producing stronger suppression of the throat-transmitted coupling. This behavior is distinct from the ordinary spatial proximity factor associated with the junction width: after the exponential SNS suppression has been divided out, a remaining power-law dependence on $\zeta_T$ or equivalently on temperature is the proposed signature of AdS$_2$ throat control. The construction is meaningful only when the matching hierarchy exists and when the infrared index is real; if $\nu_0$ becomes imaginary, the normal RN-AdS throat should instead be interpreted as unstable to charged-scalar condensation.

A sharp diagnostic is the logarithmic derivative of the residual critical current.  Define
\begin{equation}
\mathcal J_{\rm res}(\chi,T)
=
J_c(\chi,T,\ell)
\exp\!\left[\frac{\ell}{\xi_J(\chi,T)}\right]
A_J(\chi,T)^{-1}.
\label{5.16}
\end{equation}
If the residual is throat dominated, then
\begin{equation}
\frac{\partial\log\mathcal J_{\rm res}}
{\partial\log T}
\longrightarrow
2\nu_0
\qquad
(T/\Lambda_{\rm IR}\rightarrow0,\;\nu_0\in\mathbb{R}).
\label{5.17}
\end{equation}
Failure of Eq. \eqref{5.17} would not invalidate the holographic Josephson construction.  It would mean only that the charge dependence of the critical current is governed by nonuniversal ultraviolet matching, by the finite width of the throat, or by backreaction and condensate effects beyond the fixed RN-AdS treatment.

The same logic applies to the phase stiffness.  Removing the ordinary weak-link suppression gives a residual stiffness
\begin{equation}
\mathcal K_{\rm res}(\chi,T)
=
K_J(\chi,T,\ell)
\exp\!\left[\frac{\ell}{\xi_J(\chi,T)}\right]
A_K(\chi,T)^{-1}.
\label{5.18}
\end{equation}
A throat-controlled Josephson branch predicts that $\mathcal K_{\rm res}$ and $\mathcal J_{\rm res}$ carry the same leading infrared exponent, although their ultraviolet amplitudes need not coincide.  This is a useful internal check because $J_c$ probes the maximal current on the nonlinear branch, while $K_J$ probes only the small-phase response.

There is one important qualification.  When $\nu_0$ is imaginary, the RN-AdS normal throat is already unstable.  In that case Eqs. \eqref{5.14}--\eqref{5.17} are not the correct late-infrared description; the geometry and matter fields must be replaced by the condensed solution appropriate to the broken phase.  The imaginary-$\nu_0$ regime is therefore better interpreted as the onset of superconductivity than as a stable throat through which an already-formed Josephson coupling is perturbatively transmitted.

\subsection{Non-extremal and Schwarzschild-AdS limits}
The finite-density non-extremal regime is characterized by a horizon electric field but no parametrically long throat.  In this case \(\zeta_T\) is not hierarchically smaller than the radial matching scale, and the leading logarithmic part of the throat length obeys
\begin{equation}
\frac{\mathcal L_{\rm throat}^{\rm log}}{L_2}
=
\log\!\left(\frac{\zeta_{\rm UV}}{\zeta_T}\right)
=\mathcal{O}(1).
\label{5.19}
\end{equation}
The factor $\mathcal R_{\rm IR}$ is then a smooth, nonuniversal function of the background charge and temperature.  Its effect can be absorbed into the amplitudes and coherence length appearing in Eq. \eqref{5.13}.  This regime may still show substantial charge-density dependence, but that dependence is not controlled by a universal AdS$_2$ scaling exponent.

The Schwarzschild-AdS limit is still more restrictive.  Setting \(Q=0\) removes the homogeneous background charge supporting the RN-AdS geometry and gives
\begin{equation}
A_t^{\rm(bg)}=0,
\qquad
f_{\rm SAdS}(r)
=
\frac{r^2}{L^2}
-\frac{r_h^3}{L^2r},
\qquad
T_{\rm SAdS}=\frac{3r_h}{4\pi L^2}.
\label{5.20}
\end{equation}
There is no background horizon electric field, no AdS$_2$ throat, and no RN-AdS infrared scaling dimension.  The weak-link Maxwell field may still be turned on as a probe field with boundary value \(M_t(r,x)\to\mu(x)\); what vanishes in the Schwarzschild-AdS limit is the homogeneous charge sector of the geometry. The weak-link observables must therefore reduce to those of the standard holographic Josephson junction once the same boundary chemical-potential profile and scalar quantization are chosen \cite{Horowitz:2011dz}.

\begin{equation}
\begin{array}{ccl}
Q=0
&:&
\text{Schwarzschild-AdS weak link},\\[2mm]
0<\chi<1\ \text{with}\ 1-\chi^2=\mathcal{O}(1)
&:&
\text{finite-density RN-AdS weak link without throat scaling},\\[2mm]
1-\chi^2\ll1
&:&
\text{near-extremal RN-AdS weak link with possible AdS$_2$ control}.
\end{array}
\label{5.21}
\end{equation}
Only the last line gives a parametrically controlled infrared scaling region.  The proposed novelty of the present construction lies precisely in separating this last effect from the ordinary spatial proximity effect of the SNS barrier and from the smooth finite-density corrections already present away from extremality.

The consistency requirement is therefore twofold.  First, as $\chi$ is moved away from unity, the residual throat scaling should disappear continuously into smooth finite-density behavior.  Second, as $Q\rightarrow0$, the entire RN-AdS contribution must decouple, leaving the known Schwarzschild-AdS Josephson phenomenology.  These limits ensure that near-extremal throat control is an additional layer of physics rather than a replacement for the established holographic weak-link mechanism.

\section{Conclusions}
We have formulated a charged holographic weak-link construction in which the Josephson response is controlled not only by the boundary junction profile but also by the Reissner--Nordstrom-AdS charge sector.  The central point is that the black hole charge is not itself a Josephson degree of freedom: the Josephson phase difference belongs to the boundary condensate generated by the charged scalar.  The RN-AdS background instead supplies a finite-density gravitational environment in which the scalar instability, condensate profile, current response, and near-horizon infrared geometry can be varied systematically.  In this sense, the construction refines the standard holographic SNS junction by promoting the charge and near-extremality of the bulk geometry to physical control parameters for phase-sensitive transport.

The resulting observables are the current-phase relation, the critical current, the effective coherence length, and the small-phase stiffness.  The ordinary SNS interpretation requires that the central region of the inhomogeneous chemical-potential profile suppress the condensate while the exterior regions remain superconducting.  In this regime, the leading current-phase relation is expected to be sinusoidal, the critical current should decay exponentially with the junction width, and the midpoint condensate should display the corresponding half-exponent decay.  Agreement between the coherence length extracted from the critical current and that extracted from the central condensate provides a direct internal check that the configuration is genuinely operating as a holographic SNS weak link.

The RN-AdS-specific contribution appears when the finite-density black brane approaches extremality.  In that limit the near-horizon region develops an AdS$_2\times\mathbb{R}^2$ throat, and radial propagation through this throat can modify the boundary Josephson coupling by an infrared scaling factor.  Separating this radial throat contribution from the ordinary spatial proximity suppression gives a clean diagnostic of whether the critical current and phase stiffness are controlled by the near-extremal infrared dimension of the charged scalar.  Away from extremality, the same charge sector may still modify amplitudes and coherence lengths, but such dependence is smooth finite-density behavior rather than parametrically controlled throat scaling.  In the further Schwarzschild-AdS limit, the RN-AdS contribution must decouple and the construction must reduce to the established holographic Josephson phenomenology.

Several extensions are natural.  A fully backreacted treatment would determine how the condensate and weak-link current deform the geometry, especially at very low temperature where the probe approximation becomes unreliable.  A time-dependent extension would address the AC Josephson effect and the holographic implementation of a genuine voltage bias.  Adding a magnetic field would allow Fraunhofer-type interference and the study of vortex-mediated weak-link transport.  One may also compare alternative scalar quantizations, different barrier profiles, and p-wave or higher-derivative generalizations.  Finally, the exterior Josephson response developed here may be compared with condensate dynamics behind the horizon in hairy charged black holes, where Josephson-like oscillations appear as part of the interior evolution rather than as an ordinary boundary weak-link effect.

\acknowledgments
A. \"O. and R. P. would like to acknowledge networking support of the COST Action CA21106 - COSMIC WISPers in the Dark Universe: Theory, astrophysics and experiments (CosmicWISPers), the COST Action CA22113 - Fundamental challenges in theoretical physics (THEORY-CHALLENGES), the COST Action CA21136 - Adscaling observational tensions in cosmology with systematics and fundamental physics (CosmoVerse), the COST Action CA23130 - Bridging high and low energies in search of quantum gravity (BridgeQG), and the COST Action CA23115 - Relativistic Quantum Information (RQI) funded by COST (European Cooperation in Science and Technology). A. \"O. also thanks to EMU, TUBITAK, ULAKBIM (Turkiye) and SCOAP3 (Switzerland) for their support.

\bibliography{ref}

\end{document}